# Ring Laser Gyro: Noise Improves Performance

M. M. Tehrani and Shahin Soltanieh

Innova Photonics, Westlake Village, CA, USA

## Abstract

We present a systematic formulation of the performance of a ring laser gyro (RLG) under the influence of a randomized sinusoidal dither. We develop the Fokker-Planck equation of such a system and find its steady state solution that we use to find exact equations for the average output rate and its variance. The former determines the RLG's scale factor and the latter its angular random walk. We find a key parameter, called the residual lock-in rate, which affects both the scale factor and the random walk of the gyro. We show that when the noise amplitude is much larger than the residual lock-in rate the gyro rate output approaches that of an ideal gyro. However, this is at the expense of generating an angular random walk that is proportional to the residual lock-in rate. We present a formulation of the statistics of gyro output count and relate the average count and its standard deviation to gyro parameters. For example, we show that the angular random walk, in addition to being proportional to the residual lock-in rate, is also proportional to the inverse square root of the dither frequency.

## I. Introduction

The two-mode ring laser is a remarkable instrument in that it not only improved our understanding of the effects of mode competition and random processes on laser operation but also, as a gyroscope, revolutionized the inertial navigation technology.

The two counter-propagating beams circulating around the cavity can compete with each other for the photons emitted by the gain medium. The degree of competition is determined by the frequency difference between the laser operating frequency and the center of the emission line [1]-[2]. At the same time, the spontaneous emission of photons by the gain medium imparts noise on the two beams with some striking results not seen in single-mode lasers. These effects were extensively studied theoretically and experimentally [3]-[4]. For example, it was shown that for equal pump parameters and when the laser is operating at the line center where the mode competition is strongest neither of the two beams achieves full coherence no matter how far above threshold the laser is operating. And, for unequal pump parameters the stronger beam achieves coherence whereas the weaker beam does not.

The interesting physics that the mode competition exhibits in the ring laser is detrimental to its operation as a gyro. For this reason, a two-isotope neon gas mixed with helium is used as its gain medium to ensure that mode competition does not occur [5]. Thus, the two modes achieve the highest degree of coherence consistent with their pump parameters and the total cavity losses.

The behavior of a two-mode ring laser is governed by a set of four coupled nonlinear differential equations involving the intensities and phases of the two beams [6]. To simplify the set and focusing on the phases, we assume that the coupling between the two intensities is weak and they reach a steady state. The assumption reduces the set to a single nonlinear differential equation about the phase *difference* between the two counter propagating beams. Backscattering of one mode into the direction of the other that leads to the lock-in phenomenon can be included in this equation [7] and will be the starting point of our investigation in this paper.

To meet the performance needs of aircraft navigation systems, the ring laser gyro (RLG) imposed demanding requirements on its components [8]. For instance, the need for very narrow linewidth to achieve high resolution detection, required the optical cavity mirrors with total losses of under one parts per million with little change over the RLG lifetime. This, in turn, required the development of very low loss, dense and highly stable multilayer thin film technology. Also, to minimize the phase locking between the two counter-propagating beams at low rotation rates (the lock-in effect), mirror scatterings had to be extremely low. This led to the development of glass polishing techniques with sub-angstrom RMS roughness (known as Super Smooth Polishing). These technologies were used for the development of other high precision optical devices and instruments including LIGO.

Despite the best effort, RLG mirrors have always had some scattering that resulted in the lock-in effect and phase equation nonlinearities. An effective method to circumvent the lock-in effect has been to apply a dither to the gyro body. Simulations of the RLG phase equation with a sinusoidal dither has shown the existence of nonlinearities in the RLG scale factor whose magnitude depends on the dither parameters [9]. Such nonlinearities also appear at all harmonics of the dither frequency.

In Refs. [1]-[4] the effects of noise on the phases of the two counter-propagating beams were not considered. The subject was later studied in a number of papers using various mathematical techniques [10]-[11]. In these papers, expressions for the average gyro output as a function of noise parameters were derived but calculations of output variance were not presented. The latter is very significant as it relates to the gyro angular random walk (ARW), which is a key parameter for gyro performance and its alignment time.

In this paper, we present a systematic derivation of the equations governing an RLG with randomized dither. In Section II, we obtain the phase equation of the RLG under a sinusoidal dither. In Section III, we add a random noise to the resulting equation to represent the randomized dither and use the Fokker-Planck technique to develop the differential equation of the probability density of the gyro phase as a function of time. We then obtain the steady state solution to the Fokker-Planck equation in Section IV. Calculations of the average output phase and its variance are presented in sections V and VI, respectively. Although the focus of our discussion is the effects of added noise to the RLG but the formulation applies to any other random noise, including the spontaneous emissions of photons, as long as its correlation time is less than a characteristic time (Eq. 10, below),. The relationships between gyro output count and output phase are shown in Section VII. Finally, we summarize the main conclusions of this paper in Section VIII.

Although the RLG is the focus of our presentation here, the mathematical framework and the formulation are applicable to any other nonlinear oscillator subjected to a Gaussian white noise. In RLG the phase nonlinearity is represented by a sinusoidal function which can be replaced by any other function appropriate for any given nonlinear oscillator.

## II. RLG with Dither

We start with the RLG phase equation that describes the rate of change in the phase difference, $\phi$, between the two counter-propagating waves as a function of input rotation rate, $\Omega$, and the magnitude of the lock-in band $\Omega_L$:

$$\dot{\phi}(t) = \Omega - \Omega_L Sin(\phi)$$

The above equation, known as Adler equation, represents the locking behavior of two coupled oscillators when their frequencies are at or below a certain threshold [12]. In an RLG, the coupling between the two waves occurs via the scattering from the RLG mirrors or other components. The two-isotope gain medium does not make any measurable contribution to the coupling. In fact, measurements on a large number of RLGs showed that, to within the spontaneous photon noise limit, an RLG has the same lock-in band as its empty cavity counterpart [13].

We now add a sinusoidal input to the Adler equation to represent the dither motion:

$$\dot{\phi}(t) = \Omega + \Omega_d Sin(\omega_d t) - \Omega_L Sin(\phi) \tag{1}$$

where, $\Omega_d$ and $\omega_d$ are the dither peak amplitude and frequency, respectively. Time integration of (1) yields:

$$\phi(t) = \Omega t - \theta_d Cos(\omega_d t) - \Omega_L \int Sin(\phi) dt + \phi_0 \tag{2}$$

$\theta_d$ in (2) is the dither angle given by $\theta_d = \dfrac{\Omega_d}{\omega_d}$ and $\phi_0$ is the integration constant. If we use (2) for $\phi$ in the sine function of (1) and keep terms up to the first order in $\Omega_L$, we obtain:

$$\dot{\phi}(t) = \Omega + \Omega_d Sin(\omega_d t) - \Omega_L Sin[\Omega t - \theta_d Cos(\omega_d t) + \phi_0] \tag{3}$$

On the other hand,

$$Sin[\Omega t - \theta_d Cos(\omega_d t) + \phi_0] = \operatorname{Im} e^{i[\Omega t - \theta_d Cos(\omega_d t) + \phi_0]} = \operatorname{Im}[e^{i[\Omega t + \phi_0]} \times e^{-i\theta_d Cos(\omega_d t)}] \tag{4}$$

We also have [14]:

$$e^{izCos\xi} = \sum_{n=-\infty}^{\infty} i^n J_n(z) e^{in\xi} \quad , \quad J_{-n}(z) = (-1)^n J_n(z) \tag{5}$$

where $J_n(z)$ is the Bessel function of $n^{th}$ order. Application of relations (5) to (4) yields:

$$Sin[\Omega t - \theta_d Cos(\omega_d t) + \phi_0] = J_0(\theta_d)Sin(\Omega t + \phi_0) +$$

$$2\sum_{k=1}^{\infty}\left[J_{2k}(\theta_d)Cos2k(\frac{\pi}{2}+\omega_d t)\bullet Sin(\Omega t+\phi_0)+J_{2k-1}(\theta_d)Sin(2k-1)(\frac{\pi}{2}+\omega_d t)\bullet Cos(\Omega t+\phi_0)\right]$$

Due to the relatively high dither frequency (typically several hundred Hz) all terms in the summation will average to zero and we obtain:

$$\dot{\phi}(t) = \Omega + \Omega_d Sin(\omega_d t) - J_0(\theta_d)\Omega_L Sin(\Omega t + \phi_0) \tag{6}$$

Comparison of (6) and (1) indicates that the sinusoidal dither reduces the lock-in threshold from $\Omega_L$ to

$$\Omega_R \equiv J_0(\theta_d)\Omega_L \tag{7}$$

that we call Residual Lock-in Threshold. Eq. (6) also suggests that the lock-in effect may be eliminated by choosing the dither angle to coincide with any of the zeros of $J_0(\theta_d)$. However, at dither frequencies of several hundred hertz very large peak dither rates are needed to bring $\theta_d$ close to the first zero of $J_0(\theta_d)$ which occurs at $\theta_d = 2.405$. In addition, maintaining the dither angle at a precise level in all environments is a rather difficult task. Nevertheless, dithered RLGs with static lock-in rates of several degrees per second have been able to measure the earth rotation rate.

Regardless of dither parameters, in each dither cycle the RLG spends some time in the lock-in zone and, thus, it incurs some error. These errors can accumulate in time and lead to a substantial output error over the measurement time. For this reason, it has been customary to randomize the dither amplitude in each dither cycle by imparting a random signal to the dither mechanism at each dither reversal. Although dither randomization helps break up the RLG error accumulation it generates a new error called angular random walk, as will be discussed in below sections.

## III. RLG with Dither and noise

It was shown in the previous section that application of a sinusoidal dither to an RLG reduces its lock-in range by a factor of the zero order Bessel function of the dither angle. We now add a random function to (6) to represent the added noise. Depending on the dither mechanism design, the noise can be applied to dither amplitude or frequency or both. In all cases we consider the noise as an additive term to the input rate. One might argue that adding noise to dither mechanism requires a multiplicative random process (through the expression for $\Omega_R$) in the phase equation. However, the distinction between multiplicative and additive noise may not be significant for a one variable equation. For time independent $\Omega$ and $\Omega_R$ the multiplicative noise always becomes an additive noise by a simple transformation of variables [15].

Combining (6) and (7) and adding a noise term, we obtain:

$$\dot{\phi}(t) = \Omega + \Omega_d Sin(\omega_d t) - \Omega_R Sin(\phi) + \zeta(t) \tag{8}$$

where $\zeta(t)$ is the random function that represents the added noise. We assume $\zeta(t)$ to be a zero-mean $\delta$-correlated Gaussian white noise. Thus,

$$<\zeta(t)>=0 \quad \text{and} \quad <\zeta^*(t)\zeta(t')>= 2q\delta(t'-t) \tag{9}$$

$q$ in (9) is the added noise amplitude. The $\delta$ correlation of $\zeta(t)$ is an approximation. In fact, the finite bandwidth of the dither mechanism imposes a finite correlation time $\tau_{cor}$ on $\zeta(t)$. It can be shown that our approximation is valid if

$$\tau_{cor} \ll \frac{1}{\Omega_R} \tag{10}$$

which is generally the case.

In what follows, we will be dealing with various averages of $\dot{\phi}(t)$ taken over measurement times that are typically much longer than the dither period and ,hence, the sinusoidal dither term in (8) averages to zero. Thus, we have:

$$\dot{\phi}(t) = \Omega - \Omega_R Sin(\phi) + \zeta(t) \tag{11}$$

Eq. (11) is the RLG's stochastic Langevin equation which corresponds a Fokker-Planck equation for the probability density of the gyro phase. We use procedures described in [15] to transform the Langevin equation (11) with the assumptions (9) to the following Fokker-Planck equation:

$$\frac{\partial}{\partial t}P(\phi,t) = -\frac{\partial}{\partial \phi}\{[\Omega - \Omega_R Sin(\phi)]P(\phi,t)\} + q\frac{\partial^2}{\partial \phi^2}P(\phi,t) \tag{12}$$

where $P(\phi,t)$ is the probability density of gyro phase having the value $\phi$ at time $t$. The term $[\Omega - \Omega_R Sin(\phi)]$ on the right hand side of (12) is the drift vector and includes the particular phase nonlinearity of RLG. This term can be replaced by the function that represents the nonlinearity of any other sensor under consideration.

Eq. (12) can be recast as:

$$\frac{\partial}{\partial t}P(\phi,t) + \frac{\partial}{\partial \phi}G = 0 \tag{13}$$

where $G$ is the "probability current", given by:

$$G = q\left[[\Omega_0 - \Omega_{R0} Sin(\phi)]P(\phi) - \frac{\partial}{\partial \phi}P(\phi)\right] \tag{14}$$

with:

$$\Omega_0 = \frac{\Omega}{q} \quad \text{and} \quad \Omega_{R0} = \frac{\Omega_R}{q} \qquad (15)$$

introduced as dimensionless input and residual lock-in rates, respectively.

## IV. Steady State Probability Density Function

In general, and after a very short time all transient solutions of the Fokker-Planck equation die out and the system reaches a steady state characterized by:

$$\frac{\partial}{\partial t} P_s(\phi, t) = 0 \qquad (16)$$

and, Eq. (12) becomes:

$$\frac{d^2}{d\phi^2} P_s(\phi) - \frac{d}{d\phi}\left\{\left[\Omega_0 - \Omega_{R0} Sin(\phi)\right] P_s(\phi)\right\} = 0 \qquad (17)$$

Due to the fact that the system is periodic in phase with a $2\pi$ period, $P_s(\phi)$ in (17) should satisfy the following boundary conditions:

$$P_s(\phi + 2\pi) = P_s(\phi) \quad \text{and} \quad \int_0^{2\pi} P_s(\phi) d\phi = 1 \qquad (18)$$

It can be shown [16] that the solution to (17) with the boundary conditions (18) is given by:

$$P_s(\phi) = \frac{1}{N} e^{\Omega_0 \phi + \Omega_{R0} Cos(\phi)} \int_\phi^{\phi + 2\pi} e^{-\Omega_0 \phi' - \Omega_{R0} Cos(\phi')} d\phi' \qquad (19)$$

where $N$ is the normalization constant. In Appendix A, we find the expression for $N$ as:

$$N = 4\pi^2 e^{-\pi \Omega_0} \left| I_{i\Omega_0}(\Omega_{R0}) \right|^2 \qquad (20)$$

where $I_\nu(z)$ is the Modified Bessel Function of the First Kind.

## V. RLG Output Rate and Scale Factor

We can now use the steady state solution given in (19) and (20) to determine various moments of the probability distribution. In particular, we are interested in the first two moments that relate to the RLG average output rate and its variance. Taking the average of Eq. (11), we can write:

$$< \dot{\phi}(t) >= \Omega - \Omega_R < Sin(\phi) > \tag{21}$$

where:

$$< Sin(\phi) >= \int_0^{2\pi} Sin(\phi) P_s(\phi) d\phi = \frac{1}{N} \int_0^{2\pi} Sin(\phi) e^{\Omega_0 \phi + \Omega_{R0} Cos(\phi)} d\phi \int_\phi^{\phi+2\pi} e^{-\Omega_0 \phi' - \Omega_{R0} Cos(\phi')} d\phi' \tag{22}$$

Calculation of $< Sin(\phi) >$ follows the same steps we followed in calculating $< Cos(\phi) >$ shown in Appendix B (the $K$ integral). Substituting the result for $< Sin(\phi) >$ in (21), we obtain:

$$< \dot{\phi}(t) >= \Omega \left( \frac{Sinh(\pi\Omega_0)}{\pi\Omega_0} \right) \left| I_{i\Omega_0}(\Omega_{R0}) \right|^{-2} \tag{23}$$

and, the RLG Scale Factor given as:

$$\frac{< \dot{\phi}(t) >}{\Omega} = \left( \frac{Sinh(\pi\Omega_0)}{\pi\Omega_0} \right) \left| I_{i\Omega_0}(\Omega_{R0}) \right|^{-2} \tag{24}$$

To get some insight into relations (23) and (24) we use the series expansion of $\left| I_{i\Omega_0}(\Omega_{R0}) \right|^2$. We show in Appendix B that

$$\left| I_{i\Omega_0}(\Omega_{R0}) \right|^2 = \frac{Sinh(\pi\Omega_0)}{\pi\Omega_0}(1 + S_1) \tag{25}$$

where $S_1$ is a sum defined as :

$$S_1 \equiv \sum_{m=1}^{\infty} \frac{(2m)!}{(m!)^2 \prod_{k=1}^{m}(k^2 + \Omega_0^2)} \left( \frac{\Omega_{R0}}{2} \right)^{2m} \tag{26}$$

Inspection of (26) shows that successive terms in $S_1$ diminish rapidly (as $1/m^2$) for all reasonable values of $\Omega_0$ and $\Omega_{R0}$, and one has to keep the first three or four terms for any numerical evaluation.

If we use (25) and (26) in (23) the scale factor equation becomes:

$$\frac{< \dot{\phi} >}{\Omega} = \frac{1}{1 + S_1} \tag{27}$$

It can be shown that when the random input rate is much smaller than the residual lock-in rate, Eq. (27) reduces to:

$$\frac{<\dot{\phi}(t)>}{\Omega} \approx \begin{cases} 0 & \text{if} \quad \Omega_0 < \Omega_{R0} \\ \sqrt{1-(\frac{\Omega_{R0}}{\Omega_0})^2} & \text{if} \quad \Omega_0 > \Omega_{R0} \end{cases} \quad (28)$$

which is the standard scale factor equation in the absence of random input [17]. In the opposite limit, we obtain:

$$\frac{<\dot{\phi}(t)>}{\Omega} \approx 1 - \frac{\Omega_{R0}^2}{2(1+\Omega_0^2)} \quad (29)$$

In practice, the latter is the case and the second term on the right hand side of (29) represents the scale factor nonlinearity.

It is seen that the larger the noise amplitude compared to the residual lock-in rate the more the RLG output approaches that of an ideal gyro. However, this improvement is accompanied by the emergence of a new error, as we shall see below. Nevertheless, to our knowledge, RLG is the only instrument that adding noise "improves" its performance. We believe this is due to the nonlinear nature of the instrument.

## VI. RLG Output Rate Fluctuations

Having found the gyro average output we now calculate fluctuations about the average as a function of gyro parameters. It will be seen in Section VII that these fluctuations determine the gyro angular random walk. Fluctuations about the average are defined as:

$$<(\Delta\dot{\phi})^2> \equiv <(\dot{\phi}-<\dot{\phi}>)^2> = <(\dot{\phi})^2> - <\dot{\phi}>^2 \quad (30)$$

Neglecting the input noise variance, it seen that:

$$<(\dot{\phi})^2> = <[\Omega - \Omega_R Sin(\phi)]^2> \quad (31)$$

We show in Appendix B that the averaging process in (31) can be performed exactly to yield:

$$<(\dot{\phi})^2> = 2\pi\Omega G + \frac{q}{2}\left[\frac{I_{1-i\Omega_0}(\Omega_{R0})}{I_{-i\Omega_0}(\Omega_{R0})} + \frac{I_{1+i\Omega_0}(\Omega_{R0})}{I_{i\Omega_0}(\Omega_{R0})}\right] \quad (31)$$

where $G$ is the probability current defined in (14). If we substitute (31) and (23) in (30), we obtain:

$$<(\Delta\dot{\phi})^2> = 2\pi\Omega G + \frac{q}{2}\Omega_{R0}\left|I_{i\Omega_0}(\Omega_{R0})\right|^{-2}\left[I_{i\Omega_0}(\Omega_{R0})I_{1-i\Omega_0}(\Omega_{R0}) + I_{-i\Omega_0}(\Omega_{R0})I_{1+i\Omega_0}(\Omega_{R0})\right] - 4\pi^2 G^2$$
(32)

and expressing the above in terms of $S_1$ and $S_2$ yields:

$$\frac{<(\Delta\dot{\phi})^2>}{\Omega_R^2} = \left(\frac{\Omega_0}{\Omega_{R0}}\right)^2 \left(\frac{1}{1+S_1}\right)^2 \left[S_1 + \frac{\Omega_{R0}^2}{2\Omega_0^2(1+\Omega_0^2)}(1+S_1)(1+S_2)\right] \quad (33)$$

where $S_2$ is given in Appendix B, Eq.(B25), as:

$$S_2 = \sum_{m=1}^{\infty} \frac{(2m+1)!}{(m!)^2 \prod_{k=1}^{m}[(k+1)^2 + \Omega_0^2]} \left(\frac{\Omega_{R0}}{2}\right)^{2m} \quad (34)$$

Similar to $S_1$, successive terms in $S_2$ diminish rapidly (as $1/m^2$) for all reasonable values of $\Omega_0$ and $\Omega_{R0}$ one has to keep the first three or four terms for any numerical evaluation.

It is to be noted that the relative fluctuation of the gyro output rate is found from (27) and (33) as:

$$\frac{<(\Delta\dot{\phi})^2>}{<\dot{\phi}>^2} = S_1 + \frac{\Omega_{R0}^2}{2\Omega_0^2(1+\Omega_0^2)}(1+S_1)(1+S_2) \quad (35)$$

We now investigate limits of (33) and (35) when the noise amplitude is larger than the residual lock-in ($\Omega_{R0} \ll 1$)

Keeping terms up the second order in $\Omega_{R0}$ in $S_1$ and $S_2$, we obtain from (26) and (34):

$$S_1 \approx \frac{2}{1+\Omega_0^2}\left(\frac{\Omega_{R0}}{2}\right)^2 \quad \text{and} \quad S_2 \approx \frac{6}{4+\Omega_0^2}\left(\frac{\Omega_{R0}}{2}\right)^2 \quad (36)$$

whose substitution in (33) and keeping terms up the second order in $\Omega_{R0}$ yields:

$$\frac{\sigma_{\dot{\phi}}}{\Omega_R} \approx \frac{1}{\sqrt{2}}\left[1 - \frac{2}{1+\Omega_0^2}\left(\frac{\Omega_{R0}}{2}\right)^2\right] \quad \text{with} \quad \sigma_{\dot{\phi}} = \sqrt{<(\Delta\dot{\phi})^2>} \quad (37)$$

and,

$$\frac{<(\Delta\dot{\phi})^2>}{<\dot{\phi}>^2} \approx \frac{1}{2}\Omega_{R0}^2 \quad (38)$$

We notice that in this limit the relative fluctuations are independent of the input rate. For low input rate compared to the noise amplitude ($\Omega_0 \ll 1$), Eq.(37) reduces to:

$$\frac{\sigma_{\dot\phi}}{\Omega_R} \approx \frac{1}{\sqrt{2}}\left[1 - \frac{\Omega_{R0}^2}{2}\right] \tag{39}$$

and in the opposite limit of $\Omega_0 \gg 1$ we get:

$$\frac{\sigma_{\dot\phi}}{\Omega_R} \approx \frac{1}{\sqrt{2}}\left[1 - \frac{1}{2}\left(\frac{\Omega_{R0}}{\Omega_0}\right)^2\right] \tag{40}$$

.

## VII. Gyro Output Counting Statistics

So far we have studied the statistical properties of the gyro phase as functions of gyro and input parameters. In practice, the input rate information is obtained from the number of counts that the counting apparatus registers within the counting interval with each count corresponding to a change of gyro phase by $2\pi$. Thus, the number of counts, $n$, registered within the time interval $T$ can be written as:

$$n = \frac{g}{2\pi}\int_0^T \dot\phi(t)dt \tag{41}$$

where g is the gyro's *geometrical* scale factor [5] (not to be confused with the Scale Factor discussed in Section V). For convenience, we set $\frac{g}{2\pi}$ equal to one.

Eq. (41) relates the statistics of the gyro counts to those of the gyro phase. In particular, we are interested in the statistical average and variance of the number of counts. The average is given by:

$$<n> = \int_0^T <\dot\phi(t)>dt = <\dot\phi(t)>T \tag{42}$$

where due to the stationarity of the $\dot\phi$ process, $<\dot\phi(t)>$ is independent of time. If we use the scale factor equation (27) in (42) the average count for the measurement time $T$ will be:

$$<n> = \frac{\Omega T}{1+S_1} \tag{43}$$

From (41), the variance of the random process $n$ is given as:

$$<(\Delta n)^2> = \int_0^T\int_0^T <\Delta\dot\phi(t_1)\Delta\dot\phi(t_2)>dt_1 dt_2 \tag{44}$$

where the integrand is the two-time correlation function of the $\dot{\phi}$ process. It can be expressed in terms of its two-time correlation coefficient, given as:

$$\lambda(t_1, t_2) \equiv \frac{<\Delta\dot{\phi}(t_1)\Delta\dot{\phi}(t_2)>}{<\dot{\phi}>^2} \tag{45}$$

Due to the stationarity of the $\dot{\phi}$ process, $\lambda(t_1, t_2)$ is a function of the difference of the two times:

$$\lambda(t_1, t_2) = \lambda(t_2 - t_1) \tag{46}$$

Using (46) and (45) in (44), the expression for variance becomes:

$$<(\Delta n)^2> = <\dot{\phi}>^2 \int_0^T\int_0^T \lambda(t_2 - t_1) dt_1 dt_2 \tag{47}$$

The double integral in (47) can be converted to a single integral via:

$$\int_0^T\int_0^T \lambda(t_2 - t_1) dt_1 dt_2 = 2\int_0^T (T - \tau)\lambda(\tau) d\tau \tag{48}$$

whose application to (47) yields:

$$<(\Delta n)^2> = \left[<\dot{\phi}>^2 \Theta(T)\right]T \tag{49}$$

where we have defined:

$$\Theta(T) \equiv 2\int_0^T (1 - \frac{\tau}{T})\lambda(\tau) d\tau \tag{50}$$

Since our Fokker-Planck equation (12) is a first order differential equation in time, it can be shown that that the correlation coefficient $\lambda(\tau)$ is, in general, of the form:

$$\lambda(\tau) = \lambda(0) e^{-\frac{|\tau|}{T_c}} \tag{51}$$

where $T_c$ is the correlation time of the $\dot{\phi}$ process. From (45) and (51), we have:

$$\lambda(0) = \frac{<(\Delta\dot{\phi})^2>}{<\dot{\phi}>^2} \tag{52}$$

If we use (51) in (50), we get:

$$\Theta(T) = 2T_c \left[ 1 - \frac{T_c}{T} + \frac{T_c}{T} e^{-\frac{T}{T_c}} \right] \lambda(0) \tag{53}$$

and the variance equation becomes:

$$< (\Delta n)^2 > = 2TT_c \left[ 1 - \frac{T_c}{T} + \frac{T_c}{T} e^{-\frac{T}{T_c}} \right] < (\Delta \dot{\phi})^2 > \tag{54}$$

Thus the gyro count variance is equal to the gyro output rate variance multiplied by a factor that is a function of measurement time and the dither noise correlation time. The two limits of the factor are of particular interest. It can be seen from (54) that:

$$\text{when} \quad T \gg T_c \quad \Rightarrow \quad < (\Delta n)^2 > \approx 2TT_c < (\Delta \dot{\phi})^2 > \tag{55}$$

and,

$$\text{when} \quad T \ll T_c \quad \Rightarrow \quad < (\Delta n)^2 > \approx T^2 < (\Delta \dot{\phi})^2 > \tag{56}$$

From the standard deviation associated with each case we arrive at:

$$\frac{\sigma_n}{\sqrt{T}} = \sqrt{2 < (\Delta \dot{\phi})^2 > T_c} = \sqrt{2T_c} \; \sigma_{\dot{\phi}} \qquad \text{for} \quad T \gg T_c \tag{58}$$

and,

$$\frac{\sigma_n}{T} = \sqrt{< (\Delta \dot{\phi})^2 >} = \sigma_{\dot{\phi}} \qquad \text{for} \quad T \ll T_c \tag{59}$$

We notice that (58) represents a random walk coefficient of the gyro counts determined by $\sigma_{\dot{\phi}}$ and the correlation time. In this case the output count standard deviation grows with the square root of measurement time. On the other hand, (59) shows that the gyro count standard deviation grows linearly with time and represents bias fluctuations whose magnitude is proportional to $\sigma_{\dot{\phi}}$.

The treatment presented here for gyro count statistics applies to any stationary random process that is imparted to the gyro and can be described by a Fokker-Planck equation which is a first-order differential equation in time. In determining the random walk and bias fluctuations caused by each random process the $\sigma_{\dot{\phi}}$ associated with that process should be in (58) and (59).

During test and operation, the measurement time is always much larger than the random input correlation time. An exact knowledge of the relationship between the correlation time and other parameters requires solution to the time-dependent probability density, $P(\phi,t)$, of the Fokker-Planck equation (12). However, if the random signal is applied to the gyro at each dither reversal, we would expect the correlation time to be of the order of half the dither period. With this assumption, Eq. (58) can be expressed as:

$$\frac{\sigma_n}{\sqrt{T}} = \frac{\sigma_{\dot{\phi}}}{\sqrt{f_d}} \tag{60}$$

where $f_d$ is the dither frequency. Eq. (60) represents the gyro angular random walk (ARW) coefficient as a function of gyro and operational parameters. The general expressions for $\sigma_{\dot{\phi}}$ is derived from Eq.(33) with its various limits given by Eqs. (37), (39), and (40).

## VIII. Summary and Conclusions

In this paper, we have presented a systematic derivation of the effects of a randomized sinusoidal dither on the performance of an RLG. We have shown that a sinusoidal dither helps reduce the size of the RLG's locking band. However, this leads to an accumulating bias error in each dither cycle. To break the cycle, a random signal has to be added to the dither at each dither reversal. This, in turn, leads to a de-coherence of the RLG phase that leads to a new error. Depending on the length of the measurement time compared to the coherence time of the random signal, the new error can appear as a bias error or an angular random walk at the gyro output.

We obtained the above results by developing the Fokker-Planck equation of an RLG that describes the time evolution of the phase probability density vs. time as a function of the gyro and operational parameters. We obtained the steady state solution to the Fokker-Planck equation that we used to develop exact expressions for the RLG's average output and its variance. These two parameters determine the gyro scale factor and its angular random walk (ARW).

We found various approximations to the exact formulas for scale factor and ARW in different operational regimes. It turns out that the residual lock-in rate plays a key role in the RLG performance. In general, we found the scale factor is most linear and most independent of the input rate when the noise amplitude is much larger than the residual lock-in rate, At the same time, the ARW reaches a constant value that is proportional to the residual lock-in rate.

The statistics of RLG output count relate to those of its phase. We established the relationship between the two and obtained expressions for the average count and its standard deviation. The latter determines the RLG's angular random walk that also depends on the coherence time of added noise. Exact derivation of the coherence time requires the time-dependent solution of the Fokker-Planck equation that we have not attempted here. We have argued that such a coherence time can be estimated as half the dither period as the random noise is added to the dither mechanism at each dither reversal. This makes the angular random walk inversely proportional to the square root of the dither frequency.

Although we have used terminologies appropriate for mechanical implementations of the randomized dither in an RLG, but the formulation equally applies to all other techniques that modulate the gyro phase. These include intra-cavity magneto-optic and electro-optic techniques that can be implemented as stand-alone devices or built into the multi-layer stacks of one or more of the RLG mirrors. However, the intra-cavity implementations of dither lead to higher cavity optical losses and, thus, reduced gyro resolution.

Our discussion has been about the effects of dither and random noise on the (active) ring laser gyros but the results are equally applicable to (passive) ring resonator gyros. As we demonstrated in [13], a passive ring resonator gyro has a lock-in threshold of the same magnitude as an active ring laser gyro of the same size and the same level of mirror scattering. In this respect, deposited waveguide ring resonator gyros using MEMS technology may also benefit from this analysis. An implementation of dither and noise in such gyros can improve their performance.

The treatment presented here can be extended to any nonlinear oscillator under the influence of any Gaussian noise whose correlation time is less than the oscillator's characteristic time (see Eq. 10). The specific nonlinearity in ring laser gyros is through the sinusoidal function. Other forms of nonlinearities can be included in the Langevin equation (11) or the drift vector of the Fokker-Planck equation (12).

Finally, although we have discussed the performance of an RLG under the influence of a random noise that is *intentionally* added to improve its performance, the formulation and the results are equally applicable to other noises that the RLG may experience. These include internal noises due to electronics or other components and environmental noises during test or operation. Formulas presented here can be used to assess the effects of any internal and environmental random process with Gaussian distribution and whose correlation time satisfy Eq. (10) on the RLG performance. In all likelihood the environment contains a large number of uncorrelated noises with different statistics, amplitudes, and correlation times. In such situations we can invoke the Central Limit Theorem and represent the entire ensemble of environmental random processes by a single Gaussian white noise as long as there is not a dominating noise in the ensemble and all correlation times satisfy Eq. (10). If these conditions are satisfied, our formulation can be used to assess the effects of internal and environmental noises on the RLG performance.

## Appendix A: Derivation of Normalization Constant

The normalization constant, N, in $P_s(\phi)$ can be determined from:

$$\int_0^{2\pi} P_s(\phi)d\phi = 1 \tag{A1}$$

which yields:

$$N = \int_0^{2\pi} e^{\Omega_0 \phi + \Omega_{R0} Cos(\phi)} d\phi \int_\phi^{\phi+2\pi} e^{-\Omega_0 \psi - \Omega_{R0} Cos(\psi)} d\psi \tag{A2}$$

Changing the variable in the second integral from $\psi$ to $\chi = \psi - \phi$, we obtain:

$$N = \int_0^{2\pi} e^{\Omega_0 \phi + \Omega_{R0} Cos(\phi)} d\phi \int_\phi^{\phi+2\pi} e^{-\Omega_0(\chi+\phi) - \Omega_{R0} Cos(\chi+\phi)} d\chi = \int_0^{2\pi} e^{-\Omega_0 \chi} d\chi \int_0^{2\pi} e^{2\Omega_{R0} Sin(\frac{\chi}{2}) Sin(\phi + \frac{\chi}{2})} d\phi \quad (A3)$$

The second integral can be manipulated to yield:

$$\int_0^{2\pi} e^{2\Omega_{R0} Sin(\frac{\chi}{2}) Sin(\phi + \frac{\chi}{2})} d\phi = 2\int_0^\pi Cosh[2\Omega_{R0} Sin(\frac{\chi}{2}) Sin(\phi)] d\phi = 2\pi I_0[2\Omega_{R0} Sin(\frac{\chi}{2})] \quad (A4)$$

Where $I_0(z)$ is the Zero-Order Modified Bessel Function of the First Kind and we have used the relation 8.431 in [14]. Thus,

$$N = 2\pi \int_0^{2\pi} e^{-\Omega_0 \chi} I_0[2\Omega_{R0} Sin(\frac{\chi}{2})] d\chi \quad (A5)$$

With a change of variable from $\chi$ to:

$$y = \frac{1}{2}(\pi - \chi) \quad \text{for} \quad 0 < \chi < \pi \quad \text{and} \quad y = \frac{1}{2}(\chi - \pi) \quad \text{for} \quad \pi < \chi < 2\pi$$

equation (A5) becomes:

$$N = 4\pi \left[ \int_0^{\pi/2} e^{-\Omega_0(\pi - 2y)} I_0[2\Omega_{R0} Cos(y)] dy + \int_0^{\pi/2} e^{-\Omega_0(\pi + 2y)} I_0[2\Omega_{R0} Cos(y)] dy \right]$$

$$= 8\pi e^{-\pi\Omega_0} \int_0^{\pi/2} Cosh(2\Omega_0 y) I_0[2\Omega_{R0} Cos(y)] dy = 4\pi^2 e^{-\pi\Omega_0} I_{i\Omega_0}(\Omega_{R0}) \bullet I_{-i\Omega_0}(\Omega_{R0})$$

$$= 4\pi^2 e^{-\pi\Omega_0} \left| I_{i\Omega_0}(\Omega_{R0}) \right|^2 \quad (A6)$$

where we have used the relation 6.681 in [14].

# Appendix B: Calculation of the Integral in Gyro Output Fluctuations

The integral in the Equation:

$$< [\Omega - \Omega_R Sin(\phi)]^2 > = 2\pi\Omega G + q\Omega_R \int_0^{2\pi} Cos(\phi) P(\phi) d\phi \quad (B1)$$

can be calculated exactly in the following manner:

$$K \equiv \int_0^{2\pi} Cos(\phi)P(\phi)d\phi = \frac{1}{N}\int_0^{2\pi} Cos(\phi)e^{\Omega_0\phi+\Omega_{R0}Cos(\phi)}d\phi \int_\phi^{\phi+2\pi} e^{-\Omega_0\phi'-\Omega_{R0}Cos(\phi')}d\phi' \qquad (B1)$$

With a change of variable from $\phi'$ to $\chi = \phi' - \phi$, we obtain:

$$\begin{aligned}K &= \frac{1}{N}\int_0^{2\pi} Cos(\phi)e^{\Omega_0\phi+\Omega_{R0}Cos(\phi)}d\phi \int_0^{2\pi} e^{-\Omega_0(\chi+\phi)-\Omega_{R0}Cos(\chi+\phi)}d\chi \\ &= \frac{1}{N}\int_0^{2\pi} e^{-\Omega_0\chi}d\chi \int_0^{2\pi} Cos(\phi)e^{2\Omega_{R0}Sin(\chi/2)Sin(\phi+\chi/2)}d\phi\end{aligned} \qquad (B2)$$

We now define a new integral:

$$M \equiv \int_0^{2\pi} Cos(\phi)e^{aSin(\phi+\chi/2)}d\phi \qquad \text{where} \qquad a = 2\Omega_{R0}Sin(\chi/2) \qquad (B3)$$

$M$ can be written as:

$$\begin{aligned}M &= \int_0^{2\pi} Cos(\phi + \chi/2 - \chi/2)e^{aSin(\phi+\chi/2)}d\phi \\ &= Cos(\chi/2)\int_0^{2\pi} Cos(\phi+\chi/2)e^{aSin(\phi+\chi/2)}d\phi + Sin(\chi/2)\int_0^{2\pi} Sin(\phi+\chi/2)e^{aSin(\phi+\chi/2)}d\phi\end{aligned} \qquad (B4)$$

But,

$$\int_0^{2\pi} Cos(\phi+\chi/2)e^{aSin(\phi+\chi/2)}d\phi = 0 \qquad (B5)$$

and

$$\begin{aligned}\int_0^{2\pi} Sin(\phi+\chi/2)e^{aSin(\phi+\chi/2)}d\phi &= \int_{\chi/2}^{2\pi+\chi/2} Sin(\theta)e^{aSin(\theta)}d\theta = \int_0^{2\pi} Sin(\theta)e^{aSin(\theta)}d\theta \\ &= 2\int_0^\pi Sin(\theta)Sinh[aSin(\theta)]d\theta\end{aligned} \qquad (B6)$$

From relation 8.411 in [14], we have:

$$\int_0^\pi Sin(\theta)Sin[zSin(\theta)]d\theta = \pi J_1(z) \qquad (B7)$$

and choosing $z = ia$ with $Sin(ix) = iSinh(x)$, we obtain:

$$i\int_0^\pi \mathrm{Sin}(\theta)\mathrm{Sinh}[a\mathrm{Sin}(\theta)]d\theta = \pi J_1(ia) = \pi i I_1(a) \tag{B8}$$

whose substitution in (B6) yields:

$$\int_0^{2\pi} \mathrm{Sin}(\phi + \chi/2) e^{a\mathrm{Sin}(\phi+\chi/2)} d\phi = 2\pi I_1(a) \tag{B8}$$

If we now use (B8) and (B5) in (B4), the integral $M$ becomes:

$$M \equiv \int_0^{2\pi} \mathrm{Cos}(\phi) e^{a\mathrm{Sin}(\phi+\chi/2)} d\phi = 2\pi \mathrm{Sin}(\chi/2) I_1[2\Omega_{R0}\mathrm{Sin}(\chi/2)] \tag{B9}$$

whose substitution in (B2) provides:

$$\begin{aligned}K &= \frac{2\pi}{N}\int_0^{2\pi} e^{-\Omega_0\chi}\,\mathrm{Sin}(\chi/2) I_1[2\Omega_{R0}\mathrm{Sin}(\chi/2)]d\chi \\ &= \frac{2\pi}{N}\left\{\int_0^\pi e^{-\Omega_0\chi}\,\mathrm{Sin}(\chi/2) I_1[2\Omega_{R0}\mathrm{Sin}(\chi/2)]d\chi + \int_\pi^{2\pi} e^{-\Omega_0\chi}\,\mathrm{Sin}(\chi/2) I_1[2\Omega_{R0}\mathrm{Sin}(\chi/2)]d\chi\right\}\end{aligned} \tag{B10}$$

A change of variable $\chi = \pi - 2y$ in the first integral and $\chi = \pi + 2y$ in the second integral yields:

$$\begin{aligned}K &= \frac{4\pi}{N}\left\{\int_0^{\pi/2} \mathrm{Cos}(y) e^{-\pi\Omega_0+2\Omega_0 y} I_1[2\Omega_{R0}\mathrm{Cos}(y)]dy + \int_0^{\pi/2} \mathrm{Cos}(y) e^{-\pi\Omega_0-2\Omega_0 y} I_1[2\Omega_{R0}\mathrm{Cos}(y)]dy\right\} \\ &= \frac{8\pi}{N} e^{-\pi\Omega_0}\int_0^{\pi/2} \mathrm{Cos}(y)\mathrm{Cosh}(2\Omega_0 y) I_1[2\Omega_{R0}\mathrm{Cos}(y)]dy \\ &= \frac{8\pi}{N} e^{-\pi\Omega_0}\int_0^{\pi/2} \mathrm{Cos}(y)\mathrm{Cos}(2i\Omega_0 y) I_1[2\Omega_{R0}\mathrm{Cos}(y)]dy \\ &= \frac{4\pi}{N} e^{-\pi\Omega_0}\left\{\int_0^{\pi/2} \mathrm{Cos}(1-2i\Omega_0 y) I_1[2\Omega_{R0}\mathrm{Cos}(y)]dy + \int_0^{\pi/2} \mathrm{Cos}(1+2i\Omega_0 y) I_1[2\Omega_{R0}\mathrm{Cos}(y)]dy\right\} \\ &= \frac{2\pi^2}{N} e^{-\pi\Omega_0}\left[I_{i\Omega_0}(\Omega_{R0}) I_{1-i\Omega_0}(\Omega_{R0}) + I_{-i\Omega_0}(\Omega_{R0}) I_{1+i\Omega_0}(\Omega_{R0})\right]\end{aligned} \tag{B11}$$

where we have used the relation 6.681.11 in [14]. If we use the expression for $N$ in (B11), we obtain:

$$K \equiv \int_0^{2\pi} \mathrm{Cos}(\phi) P(\phi) d\phi = \frac{1}{2}\left[\frac{I_{1-i\Omega_0}(\Omega_{R0})}{I_{-i\Omega_0}(\Omega_{R0})} + \frac{I_{1+i\Omega_0}(\Omega_{R0})}{I_{i\Omega_0}(\Omega_{R0})}\right] \tag{B12}$$

and the variance of the gyro output rate becomes:

$$<(\Delta\dot{\phi})^2> = 2\pi\Omega G + \frac{q}{2}\Omega_{R0}\left|I_{i\Omega_0}(\Omega_{R0})\right|^{-2}\left[I_{i\Omega_0}(\Omega_{R0})I_{1-i\Omega_0}(\Omega_{R0}) + I_{-i\Omega_0}(\Omega_{R0})I_{1+i\Omega_0}(\Omega_{R0})\right] - 4\pi^2 G^2$$

(B13)

Using B(13), the gyro out Standard Deviation in units of $\Omega_L$ can be written as:

$$\frac{\sigma_{\dot{\phi}}}{\Omega_L} = \left(\frac{\Omega_0}{\Omega_{R0}}\right)\frac{1}{\left|I_{i\Omega_0}(\Omega_{R0})\right|}\left\{\left(\frac{Sinh(\pi\Omega_0)}{\pi\Omega_0}\right)\left[1 - \frac{Sinh(\pi\Omega_0)}{\pi\Omega_0}\left|I_{i\Omega_0}(\Omega_{R0})\right|^{-2}\right] + \frac{\Omega_{R0}}{2\Omega_0^2}\left[I_{i\Omega_0}(\Omega_{R0})I_{1-i\Omega_0}(\Omega_{R0}) + I_{-i\Omega_0}(\Omega_{R0})I_{1+i\Omega_0}(\Omega_{R0})\right]\right\}^{1/2}$$

(B14)

where we have used Eq. (14) for the probability current $G$.

We now use the relationship between $I_\nu(z)$ and $J_\nu(z)$ and the series expansion of $J_\nu(z)$ given in Eq. (7), p. 147 of [18] to obtain:

$$\left|I_{i\Omega_0}(\Omega_{R0})\right|^2 = I_{i\Omega_0}(\Omega_{R0})I_{-i\Omega_0}(\Omega_{R0}) = J_{i\Omega_0}(i\Omega_{R0})J_{-i\Omega_0}(i\Omega_{R0}) = \sum_{m=0}^{\infty}\frac{(2m)!}{(m!)^2\Gamma(m+1+i\Omega_0)\Gamma(m+1-i\Omega_0)}\left(\frac{\Omega_{R0}}{2}\right)^{2m}$$

(B15)

But,

$$\Gamma(m+1+i\Omega_0)\Gamma(m+1-i\Omega_0) = (m^2+\Omega_0^2)\left[(m-1)^2+\Omega_0^2\right]\cdots\cdots(1+\Omega_0^2)\frac{\pi\Omega_0}{Sinh(\pi\Omega_0)}$$

$$= \frac{\pi\Omega_0}{Sinh(\pi\Omega_0)}\prod_{k=1}^{m}(k^2+\Omega_0^2), \quad m=1,2,3,\ldots\ldots$$

(B16)

where we have used Eq. 6.1.31 (p.256) of [19]

For m=0 we have: $\Gamma(m+1+i\Omega_0)\Gamma(m+1-i\Omega_0) = \frac{\pi\Omega_0}{Sinh(\pi\Omega_0)}$ (B17)

Thus, using (B16) and (B17) in (B15) yields,

$$\left|I_{i\Omega_0}(\Omega_{R0})\right|^2 = \frac{Sinh(\pi\Omega_0)}{\pi\Omega_0}(1+S_1)$$

(B18)

where,

$$S_1 = \sum_{m=1}^{\infty}\frac{(2m)!}{(m!)^2\prod_{k=1}^{m}(k^2+\Omega_0^2)}\left(\frac{\Omega_{R0}}{2}\right)^{2m}$$

(B20)

Similarly, we can write:

$$I_{i\Omega_0}(\Omega_{R0})I_{1-i\Omega_0}(\Omega_{R0}) + I_{-i\Omega_0}(\Omega_{R0})I_{1+i\Omega_0}(\Omega_{R0}) = 2\mathrm{Re}[I_{-i\Omega_0}(\Omega_{R0})I_{1+i\Omega_0}(\Omega_{R0}]\qquad(B21)$$

But,

$$I_{-i\Omega_0}(\Omega_{R0})I_{1+i\Omega_0}(\Omega_{R0}) = e^{-\frac{\pi}{2}(1+i\Omega_0)i} e^{-\frac{\pi}{2}(-i\Omega_0)i} J_{1+i\Omega_0}(i\Omega_{R0})J_{-i\Omega_0}(i\Omega_{R0})$$

$$= -i\sum_{m=0}^{\infty}(-1)^m \frac{(2m+1)!}{m!(m+1)!\Gamma(m+2+i\Omega_0)\Gamma(m+1-i\Omega_0)}\left(\frac{i\Omega_{R0}}{2}\right)^{2m+1} \qquad(B22)$$

Also,

$$\Gamma(m+2+i\Omega_0)\Gamma(m+1-i\Omega_0) = (m+1+i\Omega_0)\Gamma(m+1+i\Omega_0)\Gamma(m+1-i\Omega_0)$$

$$= (m+1+i\Omega_0)\prod_{k=1}^{m}(k^2+\Omega_0^2)\frac{\pi\Omega_0}{Sinh(\pi\Omega_0)}\ ,\ m=1,2,3,.... \qquad(B23)$$

so that:

$$I_{i\Omega_0}(\Omega_{R0})I_{1-i\Omega_0}(\Omega_{R0}) + I_{-i\Omega_0}(\Omega_{R0})I_{1+i\Omega_0}(\Omega_{R0}) = \frac{\Omega_{R0}}{1+\Omega_0^2}\frac{Sinh(\pi\Omega_0)}{\pi\Omega_0}[1+S_2] \qquad(B24)$$

where we have defined:

$$S_2 = \sum_{m=1}^{\infty}\frac{(2m+1)!}{(m!)^2\prod_{k=1}^{m}[(k+1)^2+\Omega_0^2]}\left(\frac{\Omega_{R0}}{2}\right)^{2m} \qquad(B25)$$

We can now express Eq. (B14) in terms of $S_1$ and $S_2$ to obtain:

$$\frac{\sigma_{\dot\phi}}{\Omega_R} = \frac{\Omega_0}{\Omega_{R0}}\frac{1}{1+S_1}\left[S_1 + \frac{\Omega_{R0}^2}{2\Omega_0^2(1+\Omega_0^2)}(1+S_1)(1+S_2)\right]^{\frac{1}{2}} \qquad(B26)$$

$$\text{If}\quad \Omega_{R0}\ll 1\quad S_1\approx\frac{2}{1+\Omega_0^2}\left(\frac{\Omega_{R0}}{2}\right)^2 \quad\text{and}\quad S_2\approx\frac{6}{4+\Omega_0^2}\left(\frac{\Omega_{R0}}{2}\right)^2 \qquad(B27)$$

Thus, in the limit of input noise much larger than the residual lock-in rate, the expressions for gyro Scale Factor and phase standard deviation become:

$$\frac{<\dot\phi>}{\Omega}\approx 1-\frac{2}{1+\Omega_0^2}\left(\frac{\Omega_{R0}}{2}\right)^2 \qquad(B28)$$

and

$$\frac{\sigma_{\dot{\phi}}}{\Omega_R} \approx \frac{1}{\sqrt{2}}\left[1 - \frac{2}{1+\Omega_0^2}\left(\frac{\Omega_{R0}}{2}\right)^2\right] \tag{B29}$$

References

[1]. Tehrani, M.M. and Mandel, L., "Mode Coupling and Detuning in a Ring Laser." Opt. Comm. 16, 16 (1976).

[2]. Tehrani, M.M. and Mandel, L., "Mode Competition in a Ring Laser." Opt. Lett. 17 No. 69, 196 (1977).

[3]. Tehrani, M.M. and Mandel, L., "Coherence Theory of the Ring Laser," Phys. Rev. A, 179, 677 (1978).

[4]. Tehrani, M.M. and Mandel, L., "Intensity Fluctuations in a Two-Mode Ring Laser," Phys. Rev. A, 179, 694 (1978).

[5]. Aronowitz, F., in *Laser Applications,* edited by M. Ross (Academic Press, New York, 1971), Vol. 1

[6]. Aronowitz, F., Phys. Rev. 139, A635 (1965); Klimontovich, Yu. L., Landa, P. S., and Lariontsev, E.G., Zh. Eksp. Teor. Fiz. 52 , 1616 (1967) [Sov. Phys.-JETP 25, 1076 (1967)]; Privalov, V. E. and Fridrikhov, S. A., Usp. Fiz. Nauk 97 (1969) [Sov. Phys. Usp. 12, 153 (1969)]; Aronowitz, F., see Ref. 5; Aronowitz, F. Appl. Opt. 11, 2146 (1972); Menegozzi, L., and Lamb Jr., W. E., Phys. Rev. A 8, 2103 (1973); O'Bryan III, C. L. and Sargent III, M. *ibid.* 8 , 3071 (1973); Hanson, D. R. and Sargent III, M. *ibid.* 9, 466 (1974); Sargent III, M., Scully, M. O., and Lamb Jr., W. E., *Laser Physics* (Addison- Wesley, Reading, Mass., 1974)

[7]. Aronowitz, F. and Collins, R. J., Appl. Phys. Lett. 9, 55 (1966); Klimontovich, Yu. L., Landa, P. S., and Lariontsev, E.G., see Ref. 6; Hanson, D. R. and Sargent III, see Ref. 3; Aronowitz, F., in *Laser Applications,* see Ref. 5.

[8]. For a history of the laser gyro development see: Heer, C. V., "History Of The Laser Gyro", Proc. SPIE, Vol.487, Physics of Optical Ring Gyros, ( October 1984); https://doi.org/10.1117/12.943242

[9]. Huthings, T. J. and Stjern, D. C., in *Proceedings of the IEEE National Aerospace and Electronics Conference , New York.* p. 549 (1978).

[10]. Cresser, J. D., Louisell, W. H., Meyster, P., Schleich, W. and Scully, M. O., Phys. Rev. A 25, 2214 (1982); Cresser, J. D., Hammons, D., Louisell, W. H., Meyster, P. and Risken, H., Phys. Rev. A 25, 2226 (1982); Cresser, J. D. Phys. Rev. A 26, 398 (1982); Schleich, W., Cha, C.

S., and Cresser, J. D., Phys. Rev. A <u>29</u>, 230 (1984); Schleich, W. and Dobiasch, P., Opt. Commun. <u>52</u>, 63 (1984).

[11]. Bambini, A. and Stenholm, S., Opt. Commun. <u>49</u>, 269 (1984); Phys. Rev. A <u>31</u>, 329 (1985); Phys. Rev. A <u>31</u>, 3741 (1985); J. Opt. Soc. Am. B <u>4</u>, 148 (1987).

[12]. Adler, R., Proc. IRE, <u>34</u>, 351 (1946). See also: Aronowitz, F. and Collins, R. J. in Ref. [7].

[13]. Tehrani, M. M. "Equivalence of Passive and Active Lock-In Rates in Ring Laser Gyros". Proc. Stuttgart Conference on Gyro Technology, Sept. 1998.

[14]. Gradshteyn, I.S. and Ryzhik, I.M., "*Table of Integrals, Series, and Products*", Corrected and Enlarged Edition. Academic Press, 1980.

[15]. Risken, H., *The Fokker-Planck Equation,* Second Ed., Springer-Verlag, Berlin, Heidelberg, NewYork (1989).

[16]. Stratonovich, R. L., *Topics in the Theory of Random Noise,* Vol. II, Chapter 9, Gordon and Breach, New York (1967).

[17]. Tehrani, M. M. and Soltanieh, S., <u>IEEE Sensors Journal</u>, Volume:19, <u>Issue: 22</u> , Nov.15, 2019). http://dx.doi.org/10.1109/JSEN.2019.2933183

[18]. Watson, G. N., *A Treatise on the Theory of Bessel Functions,* Cambridge University Press (1922).

[19]. Abramowitz, M. and Stegun, I. A., Editors, *Handbook of Mathematical Functions.* National Bureau of Standards, Applied Mathematics Series. <u>55</u> (1964).

**Biographies**

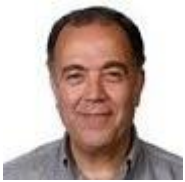

Mohammad M. Tehrani received his BS degree in physics from University of Tehran, Tehran, Iran and his MS and Ph.D in physics from the University of Rochester, Rochester, NY. As part of his Ph.D thesis, he developed the Coherence Theory of Ring Lasers and conducted experiments to verify the effects of spontaneous emission of photons and mode competition in ring lasers. He joined Honeywell Systems and Research Center in 1977 where he helped the development of ring laser and fiber optic gyros. In 2001, he founded LighTap to develop tunable devices for fiber optic networks. In 2004 he joined The Aerospace Corp. where he was a Review Team member of NASA projects specializing in space instruments. He also worked on the development of coupled GPS/IMU systems. In 2016, he co-founded Innova Photonics to develop optical instruments for biomedical applications focusing on early detection of cancer cells. He is a Senior Member of IEEE and Optical Society of America with 20 publications in open literature and nine patents and patent applications.

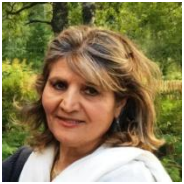

Shahin Soltanieh received her BS in physics for the University of Teharn, Tehran, Iran and her MS degree in physics from the University of Rochester, Rochester, NY. With Prof. Emil Wolf, she developed the dyadic theory of electromagnetic waves for formulation of angular correlation functions. She worked as a ring laser gyro production engineer at Northrop Grumman. In 2001, she co-founded LighTap and worked on the development and production of tunable devices for fiber optic networks. She is the co-founder of Innova Photonics in charge of instrument design. She is a Senior Member of IEEE.